\documentclass{article}

\usepackage[preprint]{neurips_2020}

\usepackage[utf8]{inputenc} 
\usepackage[T1]{fontenc}    
\usepackage{hyperref}       
\usepackage{url}            
\usepackage{booktabs}       
\usepackage{amsfonts}       
\usepackage{nicefrac}       
\usepackage{microtype}      
\usepackage{wrapfig}
\usepackage{color}
\usepackage{here}

\usepackage{graphicx}
\usepackage{subfigure}

\makeatletter
\newcommand{\figcaption}[1]{\def\@captype{figure}\caption{#1}}
\newcommand{\tblcaption}[1]{\def\@captype{table}\caption{#1}}
\makeatother

\usepackage{amsmath, amssymb,mathrsfs} 

\title{Neural Schr\"{o}dinger Equation:\\
		Physical law as Neural Network}

\author{%
Mitsumasa Nakajima, Kenji Tanaka, Toshikazu Hashimoto\\
NTT Device Technology Labs., Kanagawa, Japan \\
\texttt{mitsumasa.nakajima.wc@hco.ntt.co.jp}\\}

\begin{document}

\maketitle

\begin{abstract}
We show a new family of neural networks based on the Schr\"{o}dinger equation (SE-NET). In this analogy, the trainable weights of the neural networks correspond to the physical quantities of the Schr\"{o}dinger equation. These physical quantities can be trained using the complex-valued adjoint method. Since the propagation of the SE-NET can be described by the evolution of physical systems, its outputs can be computed by using a physical solver. As a demonstration, we implemented the SE-NET using the finite difference method. The trained network is transferable to actual optical systems. Based on this concept, we show a numerical demonstration of end-to-end machine learning with an optical frontend. Our results extend the application field of machine learning to hybrid physical-digital optimizations.
\end{abstract}
\section{Introduction}
Deep neural networks (DNNs) have a remarkable ability to learn and generalize from data. This feature enables excellent capabilities in a wide range of applications, such as image processing, natural language processing, and robot operation\cite{lecun2015deeplearning,graves2013speech,silver2016mastering,journals/corr/LillicrapHPHETS15}. Recent studies have shown that it is possible to model the residual layer of ResNet as a continuous ordinary differential equation (ODE) network \cite{journals/corr/abs-1905-10994} or by partial differential equations (PDEs). The ODEs and PDEs are commonly used to simulate physical systems. Therefore, the previous work on ODE-Nets suggests that the evolution equations used for actual physical systems can be considered as a family of neural networks. 

In this study, we propose a new building block for neural networks based on the Schr\"{o}dinger equation (SE-NET) as depicted in Fig. 1(a). The Schr\"{o}dinger equation is used for modeling quantum phenomena and also commonly used for analyzing lightwave transmission for the design of analog optical waveguides \cite{wave-theory, soloton}. As the Schr\"{o}dinger equation is related to various physics (see Fig. A1), the discussion in this paper  applies to various physical laws, including those in thermodynamics, electromagnetism, and mechanics . Our contributions in this paper are as follows:

\textbf{Network model augmentation to physical law: } We extend continuous representations of neural networks to physical law (the Schr\"{o}dinger equation). The relationship between the manipulation of the physical quantity (potential field) and the neural network is revealed. We show that the SE-NET can be trained by using the complex-valued adjoint method . In this framework, the learning targets are no longer  only the weights of DNNs but also actual physical quantities.

\textbf{Computational physics for neural network solver: }Thanks to the model augmentation, a physical simulator can be used as a neural network solver. We show that the SE-NET can be efficiently computed by using the Cranck-Nicolson finite difference method (FDM). To prevent exploding and vanishing gradients, we also introduce a phase-only optimization method that was originally developed for topology optimization of optics. 

\textbf{Physical data processing: }The SE-NET describes actual physical behaviors, such as beam propagation on an optical waveguide. The training means the "design" of physical structure, and the designed structure is physically transferable. We can train the physical structure as a black box on a machine learning (ML) platform such Pytorch.  A physical neural network provides the chance to access such natural data.

\textbf{End-to-end hybrid physical digital machine learning: } In general, much information is lost during measurement: e.g. camera images lose depth (phase), polarization, and wavelength information through measurement. By using our framework, we can jointly optimize the physical structure and digital DNNs under the same cost function. All gradients can be computed by the chain rule and adjoint. This enables end-to-end ML with a physical structure manipulating the natural information. As an example, we demonstrate a data-driven end-to-end design of an ultracompact optical spectrometer.

In the following section, we demonstrate the above-mentioned features of the SE-NET. Note that our discussion basically assumes classical (not quantum) systems. Thus, the discussion does not pertain to quantum ML systems, but there are several similarities between our learning scheme and quantum systems as described in Section 6. 
\begin{figure*}[t]
\vskip -0.5cm
\begin{center}
\centerline{\includegraphics[width=12cm]{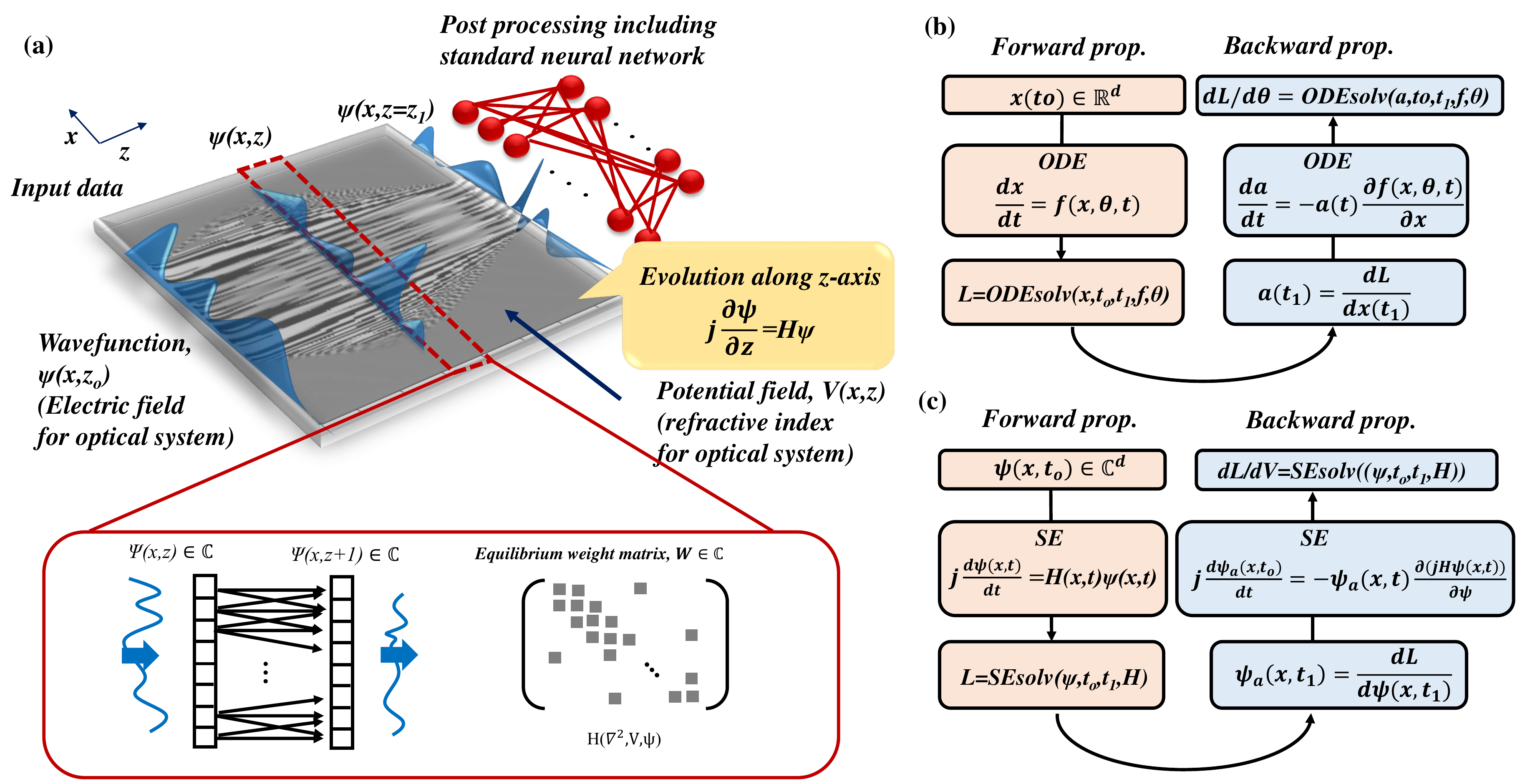}}
\caption{(a) Concept of neural SE. Diagram of (a) Neural ODE and (b) Neural SE.}
\label{fig2}
\end{center}
\vskip -0.4in
\end{figure*}
\section{Neural Schr\"{o}dinger equation}
\label{Neural Schodinger equation}
\subsection{Theory}
\textbf{Neural ODEs: } First, we describe the continuous representation of neural networks. Hidden state $x(L)\in \mathbb{R}$ for a ResNet is represented by $x(L+1)=x(L)+f[x(L),\theta]$, where $L \in \mathbb{N}$ denotes the layer number, $f$ denotes the nonlinear function, and $\theta \in \mathbb{R}$ is a trainable weight. It has been shown that a residual block can be interpreted as a discrete approximation of an ODE by setting the discretization step to one. When the discretization step approaches zero, it yields a family of neural networks, which are called neural ODEs \cite{conf/nips/ChenRBD18}. A block diagram of the processing in an ODE-Net is shown in Fig. 1(b). Formally, in a neural ODE, the relation between the input and output is characterized by
\begin{equation} \frac{dx(t)}{dt}=f[x(t),\theta,t].\end{equation}
From the given input $x(t_{o})$, the output $x(t_{1})$ can be computed by solving the ODE in (1). The backward propagation of neural ODE-Nets can also be computed with the standard ODE solver by considering their adjoint. 

\textbf{Neural Schr\"{o}dinger equation: } Next, we consider the time-dependent one-dimensional Schr\"{o}dinger equation in the neural ODE framework. A block diagram of the processing in the SE-NET is shown in Fig. 1(c). The Schr\"{o}dinger equation is a complex-valued second-order hyperbolic equation described as 
\begin{equation} j\hbar\frac{\partial\psi(x,t)}{\partial t}=\mathcal{H}\psi(x,t).\end{equation}
where $\psi(r,t)\in\mathbb{C}$ is a wave function, $t$ is time, $x\in\mathbb{R}$ is a position vector, $j$ is an imaginary unit, $\hbar$ is Dirac's constant, and $\mathcal{H}$ is the Hamiltonian of considering systems. For example, the Hamiltonian for a particle under the potential field $V(x,t)\in\mathbb{C}$ is described as follows. 
\begin{equation} \mathcal{H_{L}}=-\frac{\hbar^{2}}{2m}\frac{\partial^{2}}{\partial x^{2}}+V(x,t),\end{equation}
where $m$ denotes the mass of the particle. As discussed in the appendix and \cite{PDE-net}, one-dimensional kernel filter $K(\theta)$ in the convolutional neural network (CNN) is described as 
\begin{equation} K(\theta)=\alpha_{1}(\theta)+\frac{\partial}{\partial x}\alpha_{2}(\theta)+\frac{\partial^{2}}{\partial x^{2}}\alpha_{3}(\theta).\end{equation}
By comparing Eqs. (3) and (4), we see that linear Schr\"{o}dinger equation [Eq. (2)] corresponds to the complex-valued evolution of the ODE-Net in Eq. (1) without a nonlinear function. In this analogy, the tuning of the potential field corresponds to the change in the weights in the kernel filter. The above discussion can be extended to the three-dimensional case. Then, the dimensions of the corresponding kernel filter are expanded. The Schr\"{o}dinger equation with linear Hamiltonian does not include nonlinear conversion. Therefore, we als consider a nonlinear Hamiltonian such as $\mathcal{H_{NL}}=(\mathcal{H_{L}}+g|\psi(x,t)|^{2})$, where $g$ is a nonlinear constant and $|\cdot|$ denotes an absolute value. This equation is commonly used for the analysis of nonlinear optical phenomena and quantum chaos \cite{soloton}. Note that linear Hamiltonia is also valuable for considering actual implementation to physical SE-NET. 

As discussed above, the time evolution of the Schr\"{o}dinger equation is considered as a special type of neural network, which we call the SE-NET. In this study, we simply deal with the Shr\"{o}dinger equation as a classical complex-valued wave equation; that is,  $\psi(x,t)$ in the Schr\"{o}dinger equation simply represents the complex-valued amplitude of a transmitted wave (not the probability of the existence of a particle).

\textbf{Training by adjoint method: }As the building block of SE-NET is no longer a simple weight matrix, it seems difficult to apply the standard back-propagation method. However, as pointed out in \cite{conf/nips/ChenRBD18} and also described in the past\cite{lecun1, bp-rc}, the backpropagation of an ODE-based continuous network is equivalent to an adjoint sensitivity analysis. The adjoint method is typically used to optimize physical systems, which is called inverse design or topology optimization \cite{meta,adjointsip}. 

Here, we consider optimizing loss function $L$, whose input is the result of SE-NET outputs. To optimize $L$, we consider how the gradients of the loss depends on the hidden state $\psi(x,t)$. This quantity is called the adjoint $\psi_{a}(x,t)$, which is defined as $ \psi_{a}(x,t)\equiv \frac{\partial L}{\partial \psi(x,t)}$
Its dynamics are given by :

\begin{equation}\begin{split} \frac{\partial \psi_{a}(x,t)}{\partial t}=\psi_{a}^{*}(x,t)\left(j\mathcal{H}-\frac{\partial \mathcal{H}}{\partial \psi(x,t)}\right)\psi(x,t)
\end{split}\end{equation}
\begin{equation} \frac{\partial L}{\partial V(x_{o},t_{o})}=-j\psi_{a}^{*}(x_{o},t_{o})\frac{\partial \mathcal{H}}{\partial V(x_{o},t_{o})}\psi(x_{o},t_{o})
\end{equation}
which can be thought of as the instantaneous analog of the chain rule. From (2) and (6), we derive the following gradients for $V$:
\begin{equation} \frac{\partial L}{\partial V_{real}(x_{o},t_{o})}=Im[\psi_{a}^{*}(x_{o},t_{o})\psi(x_{o},t_{o})]\end{equation}
\begin{equation} \frac{\partial L}{\partial V_{imag}(x_{o},t_{o})}=Re[\psi_{a}^{*}(x_{o},t_{o})\psi(x_{o},t_{o})]\end{equation}

where $V_{real}(x_{o},t_{o})$ and $V_{imag}(x_{o},t_{o})$ are real and imaginary part of potential fields at each ($x_{o},t_{o})$. As shown in Eqs. (7) and (8), the SE-NET can be trained by optimizing the local potential field $V(x,t)$. The gradients can be computed by evaluating forward propagated field $\psi(x,t)$ and backpropagated adjoint field $\psi_{a}(x,t)$. As can be seen in Eq. (5), we can solve the backpropagated wave by using almost the same propagation rules: for linear Hamiltonian [$(\partial \mathcal{H}/\partial \psi)=0$], in particular, the backpropagation equation is mirrors the forward propagation one. The same training rule can be introduced by considering the calculus of variations as described in \cite{WFM1}. 

\textbf{Physical simulator as SE-NET solver: } Several techniques for efficiently solving Eqs.(2) have been widely studied \cite{BPM_text, chang1999difference}. We can use these methods as a SE-NET solver. The algorism based on fast Fourier transformation (FFT) is simple but it requires $O(n_{x}log(n_{x}))$ complexity where $n_{x}$ is number of grids along $x$-axis. Here, we solve this equation using a finite-difference (FDM) method which requires only $O(n_{x})$ complexity. There are various ways to implement the FDM. As also discussed in the original ODE-paper, the choice of solver is important because the numerical stability is different in each solver. Explicit methods such as the Euler method are simple and easy to implement. However, their solutions become unstable when $\Delta x^{2}/\Delta t>1/2$, where $\Delta x$ and $\Delta t$ are grid spacings along the $x$- and $t$-axes. In addition, we need fine computational meshes to obtain  accurate solutions because their numerical error is relatively large: $[O(\Delta x^{2})+O(\Delta t)]$. Here, we employ the Crank-Nicolson method to solve FD-BPM  because its solutions are stable and its numerical error is small, $[O(\Delta x^{2})+O(\Delta t^{2})]$. The details of the Crank-Nicolson method are described in the appendix. 

\subsection{Application to optics}
\textbf{Schr\"{o}dinger equation in optics: } It is known that lightwave propagation in an optical waveguide with paraxial ray approximation is described by the Schr\"{o}dinger equation. Therefore, we can consider the SE-NET as a way of optical beam transmission. In this analogy, the training of the SE-NET means the "design" of the waveguides. Thus, the trained network is transferable to actual optical systems. Here, we consider light propagation in a medium with a refractive index distribution for the emulation and training of the SE-NET. For optical systems, we can derive the Schr\"{o}dinger equation as follows by considering time-independent two-dimensional ($x$-$z$) systems with slowly varying envelope approximation \cite{wave-theory}: 
\begin{equation} j\frac{\partial\psi(x,z)}{\partial z}=\mathcal{H}\psi(x,z).\end{equation}
where $\psi(x,z)$ is an electric field. For the standard linear optical system, Hamiltonian $\mathcal{H_{L}}$ is described as 
\begin{equation} \mathcal{H_{L}}=\frac{1}{2kn_{r}}\left(-\frac{\partial^{2}}{\partial x^{2}}+k^{2}\Delta n(x,z)\right)\end{equation}
where $k$ is wavenumber. $\Delta n$ is defined as $\Delta n \equiv (n-n_{r})$, where $n_{r}$ is a reference refractive index $n$ is a complex valued local refractive index. The real part of the refractive index, $n_{real}$, denotes the phase shift of the wave, and the imaginary part $n_{imag}$ contributes the loss or gain of the system. By comparing with the Schr\"{o}dinger equation in Eq.(2), it is clear that the first term of Eq.(10) corresponds to momentum and second term corresponds to the local potential. The nonlinear conversions can be introduced by considering the approach described in section 2.2 (see Fig. 2). Thus, the $z$-axis evolution of Eq. (9) is also considered as forward propagation in a neural network.

By considering adjoint $\psi_{a}(x,z)\equiv \frac{\partial L}{\partial \psi(x,z)}$, we can derive the update rule  in the same way as in Eqs. (9) and (10). This means that the SE-NET can be trained by optimizing the local $n(x,z)$. We can solve both forward propagation and backpropagation by using a simulation method for optical systems. 

\textbf{Phase-only training:} We have two parameters at each position, ($x_{o}, z_{o}$). However, $n_{imag}$ means the gain or loss of the transmitting medium. ODE-based dynamical systems are known to be unstable when the gain of the weight matrix becomes large, which corresponds to gradient explosion. In addition, system loss significantly affects the signal-to-noise-ratio of the output signal, which degrades system performance. 
To avoid the above issue, we consider the phase-only update rule for the local potential $n$, which was developed for the inverse design of optical devices \cite{WFM1,WFM2}. This method is called the wave-front-matching (WFM) method. In the WFM method, $n_{imag}$ and $\partial L/\partial n_{imag}$ are fixed at zero. We only update $n_{real}$, which means that the update is executed to match the wavefront between the forward propagating wave and backpropagating wave. By using this method, the system automatically maintains unitarity. The weight matrix derived from the local refractive index is a Hermitian matrix. Thus, we can maintain model stability. This means that the system is always stable and that the law of the conservation of energy is satisfied. Therefore, the physical system has no principle loss or energy consumption for solving SE-NETs . This framework is considered as a specific expression of unitary neural network \cite{uRNN}.

\textbf{Manipulate lightwave information: } Lightwaves have inherent parallelism with respect to phase, space, and wavelength. These data are lost when intensity is detected  by a photodetector (PD). The optical SE-NET can process these data via $n$. The phase and space information can be directly tuned by $n_{real}(x,z)$. Since the amount of phase modulation tuned by  $n$ differs for each wavelength, the wavelength information can be processed by using the standard SE-NET [eqs. (9) and (10)]. Thus, we can train the physical structure by considering the wavelength. 

\textbf{Optics on Pytorch: } We implement the optical SE-NET using finite difference beam propagation method (FD-BPM) on Pytorch. Fig. 2(a) shows a training example, which is an optical 1:2 splitter design. The training example of wavelength splitter is also shown in Fig. 2(b). In our framework, we treat the wavelength information as the data in each minibatch. Although this framework is almost the same as topology optimization or inverse design \cite{WFM1}, we can optimize the physical structure as a black box with few prior physical knowledge on Pytorch. The additional information for experiments is described in appendix.

\textbf{End-to-end hybrid physical digital ML: } Thanks to Pytorch implementation, we can jointly optimize the physical structure and digital DNNs under the same cost function. All gradients can be computed by the chain rule and adjoint thanks to the \textit{autograd} function on Pytorch. This enables the end-to-end ML with a physical structure. The algorithm is shown in the appendix.

\begin{figure*}[t]
\vskip 0.1in
\begin{center}
\centerline{\includegraphics[width=12cm]{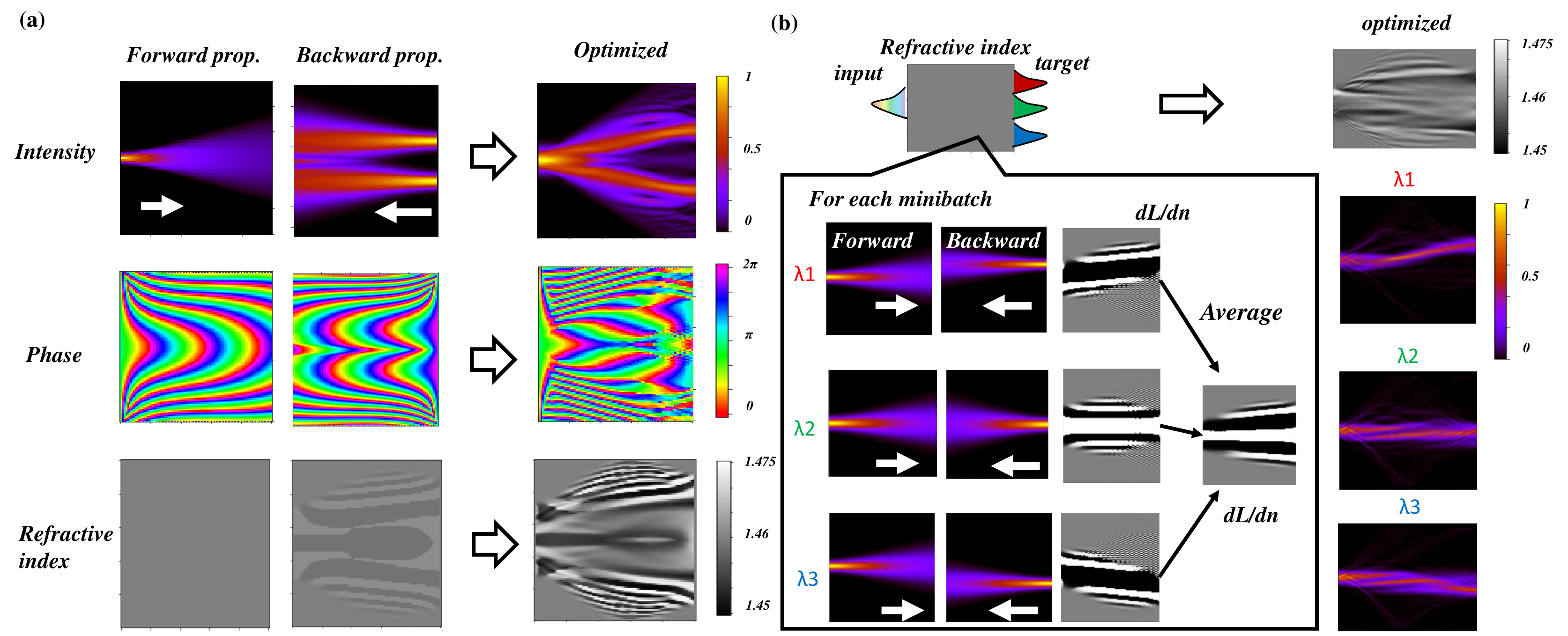}}
\caption{Learning example using physical solver called the FD-BPM on Pytorch for (a) 1:2 power splitter and (b) wavelength demultiplexer.}
\label{fig2}
\end{center}
\vskip -0.4in
\end{figure*}

\section{Experimental results}
To examine above described training framework, we validated the SE-NET using some datasets. To compute the $z$-axis evolution of SE-NET numerically, we implemented the FD-BPM solver with the Crank-Nicholson method on the Pytorch framework. To train the weights, we used a standard Adam optimizer with a learning rate of 0.001. For the training, we set  the size of the calculation grid along $x$- and $z$-axis to 1.0 $\mu$m. The reference refractive index $n_{r}$ to 1.45. These hyperparameter values for the SE-NET are based on the standard silica-based optical circuits for telecom applications.

\textbf{Image classification:} At first, we confirmed the performance of SE-NET itself by neglecting the physical implementation. We trained SE-NETs to classify the MNIST dataset of handwritten digits \cite{MNIST}. The data set consists of 60,000 labeled digital images with size $28\times28$ showing hand written digits from 0 to 9. For the training, the optimizer steps were computed using mini-batches consisting of 128 randomly chosen examples from the 60,000 training data. The input wavelength was set to $(\lambda=2\pi/k)$ to 1.55 $\mu $m. The trained models for our experiment are shown in Fig. 3(a) and (b). The architecture in Fig. 3(a) is named CNN-SE-NET. It has a CNN-based down sampling layer containing three two-dimensional convolution units, which is the same as in the pre-filtering method used in the previously reported ODE-Net model \cite{conf/nips/ChenRBD18}. To neglect the effect of the CNN layer, we also trained the model with SNN-based down sampling, named SE2-NET. Calculation regions are also shown in Fig. 3(a) and (b). We consider two types of nonlinear conversion as shown in Fig. 3(c) and (d) in this task. First one uses typical nonlinear Hamiltonian with $g=0.1$. We decided this value using a more simple task (see Appendix). The other one employs the nonlinear activation after propagation of SE-NET as shown in Fig. 3(d). This system can be physically devised by inserting the nonlinear material discretely. At the input edge of the SE-NETs, the input data are aligned to the $x$-axis and mapped to the real part of $\psi(x,z_{0})$. At the output edge of SE-NETs, the real and imaginary parts of $\psi(x,z=z_{1})$ are independently treated. They go to the full connection layer. The experimental results for test error are shown in Table I. As can be seen in this figure, the performances of the SE-NETs are on the same order of magnitude as the best performance found in the literature.  Although SE2-NET achieved 1.33\% accuracy, the CNN-SE-NET model showed much better performance, suggesting that CNN-based down sampling is superior to SNN-based down sampling. We think this difference is due to the one-dimensional feature of our tried model as described in Eq.(4). Thus, it will be improved when we consider more highly dimensional SE-NET model including time or $y$-axis transmission. The experiment for the larger dataset is also shown in appendix.
\begin{figure}[H]
\vskip -0.5cm
\begin{center}
\centerline{\includegraphics[width=12.0cm]{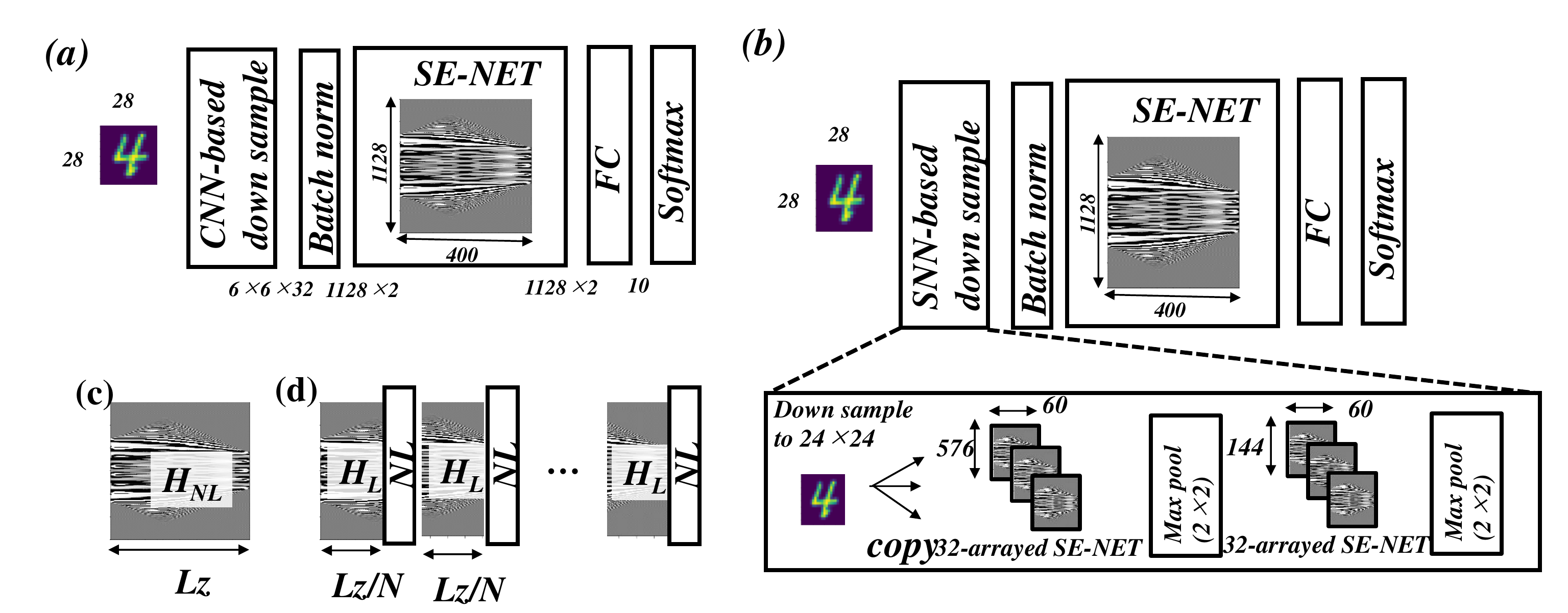}}
\caption{Trained model for MNIST task with (a) CNN- (b) SE-NET-based down sampling layers.}
\label{icml-historical}
\end{center}
\vskip -1cm
\end{figure}

\begin{table}[H]
\caption{Test error for MNIST data sets.}
\label{sample-table}
\vskip -1cm
\begin{center}
\scalebox{0.9}{
\begin{small}
\begin{sc}
\begin{tabular}{ccc||ccc}\hline
\toprule
 &CNN-SE-NET & SE2-NET & ODE-Net & LesNet \\
 Hamiltonian & (this work)  &(this work) & \cite{conf/nips/ChenRBD18}  & \cite{lenet1} \\
\midrule
\midrule
Linear(N=1)  & $0.77\%$ & $ 1.33\%$ \\
Linear(N=5) & $0.63\%$ & $ 1.05\%$ &  $0.42\%$ & $1.1\%$ \\
Nonlinear(g=0.1)  & $0.69\%$ & $ 1.02\%$ \\

\bottomrule
\end{tabular}
\end{sc}
\end{small}
}
\end{center}
\vskip -0.5cm
\end{table}
\textbf{End-to-end compressed sensing spectroscopy: }Next, we demonstrate possible physical implementation with the joint end-to-end machine learning scheme. We tried to design an ultra-compact optical spectrometer. As the photodetector lost the wavelength information by the measurement, the standard spectrometer is designed to receive each wavelength independently as shown in Fig. 4(a). It requires bulk gratings and long optical paths to separate the wavelength spatially. For the miniaturization, compressive sensing (CS) spectroscopy is attracting much attention \cite{CS4,CS5,CS1,CS2,CS3}. In this method, spectra $y\in \mathbb{R}^{N\times1}$ are reconstructed from a few detector signals $x\in \mathbb{R}^{N\times1}$ with a known relationship $x = Ty$, where $T\in \mathbb{R}^{M\times N}$ is the detection matrix of the optical structure. It was reported that  the deep learning is a powerful tool for reconstruction \cite{CS1,CS2,CS3}. Each row of $T$ represents the spectrum transmission to each detector. In general, the optical implementation of T requires complex optics and specialized knowledge for the design. 

Here, we consider the system shown in Fig. 4(b). We implement $T$ using the SE-NET, which enables simultaneous optimization of $T$ and the DNN for reconstruction. To evaluate the performance, we used 12,000 synthetic spectral datasets with wavelengths of 1000–1500 nm using  Gaussian distribution functions. The data were randomly divided into 10,000 training examples, and 2,000 examples were used for validation. The input data are represented as a complex-valued Gaussian beam with a beam width of 100 $\mu$m. The analyzed region was set to $500\times500 \mu$m$^{2}$. The 25 detectors were set on the output edge of the waveguide. We  used a linear Hamiltonian by considering easy transfer to the optics. For the post processing, we used a residual layer with a one-dimensional convolution filter, which has almost the same structure as the filter in \cite{CS2} (see the appendix for more details). Figure 4(c) and (d) shows the validation loss and peak signal to noise ratio (PSNR) with and without SE-NET updates. The result is the average of 5 examines. After the training of 20 epoch, the validation PSNR for the model with the SE-NET reached 41.7 dB, which is far superior to the PSNR without SE-NET updating (25.9 dB). Fig. 4(e) shows the two examples of reconstructed spectra. Figure 4(f) shows the evolution of refractive index under the training. As can be seen, the structure is optimized from the flat initial state, suggesting that we can optimize the physical structure as a black box with a little prior physical knowledge . These results suggest that the effectiveness of the hybrid physical digital ML.

\begin{figure*}[t]
\vskip -0.1in
\begin{center}
\centerline{\includegraphics[width=5 in]{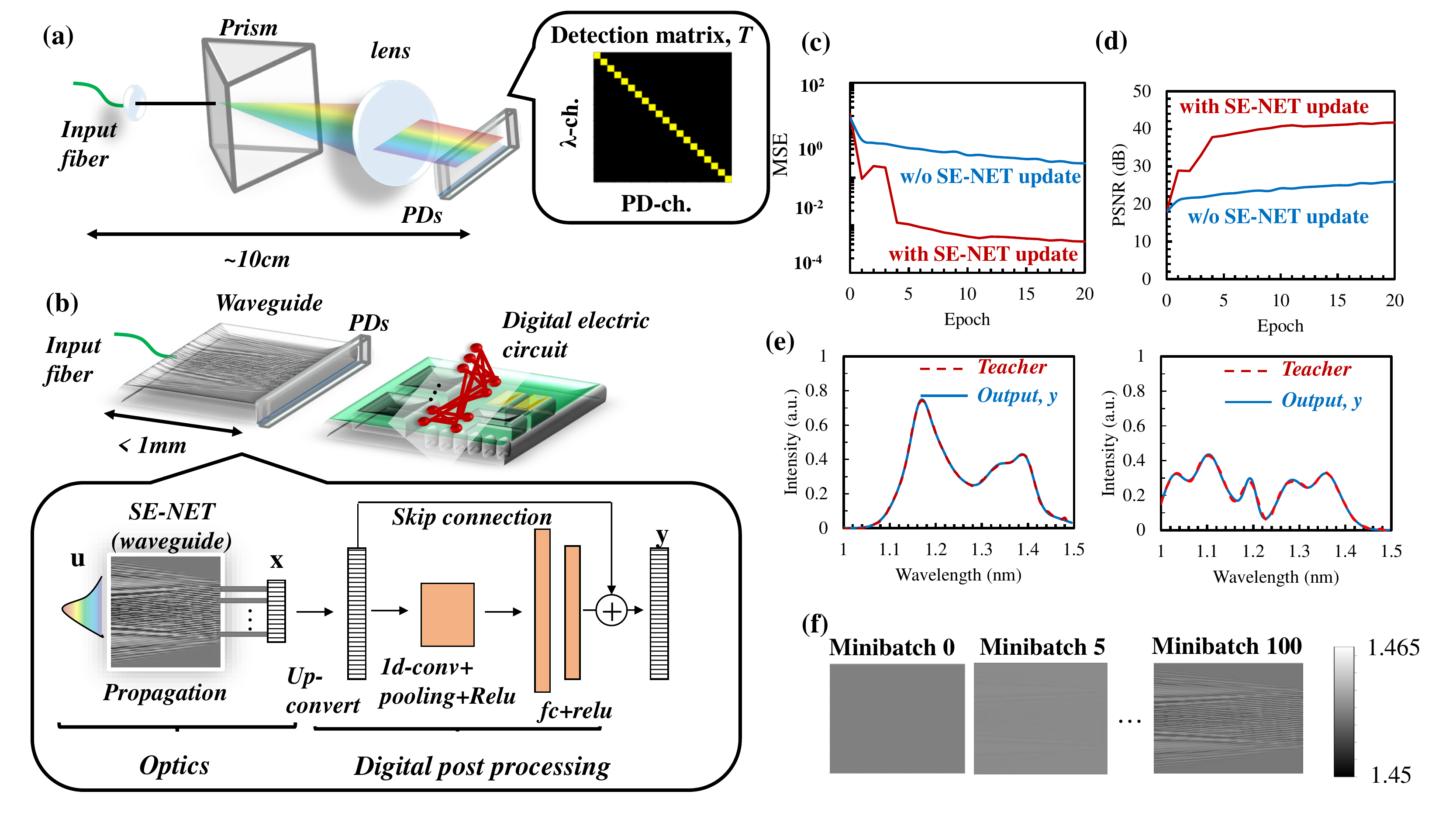}}
\caption{Schematic of (a) standard spectrometer and (b) considered CS sepectrometer. (c) loss and (d) PSNR as a function of minibatch. (e) Typical reconstructed spectra and (f) refractive index for each minibatch state}
\label{fig2}
\end{center}
\vskip -0.4in
\end{figure*}

\section{Discussion and outlook}
\textbf{Augmentablity to other wave dynamics: }We summarize the correspondence between physics and neural network in Figure A.1. The real-valued repressentation of the Schr\"{o}dinger equation corresponds to the diffusioin equation. Thus, our work could be extended to the thermal conduction, reaction-diffusion system, and fluid systems. The Schr\"{o}dinger equationis derived from the scalar wave equation. Our work could thus be extended to complex-valued propagation of, for instance, acoustic waves. As is well known, classical limitation (i.e. $\hbar\rightarrow0$ in quantum systems or paraxial approximation in optics) corresponds to the Hamilton-Jacobi equation. Therefore, our discussion can be extended to members of its family, such as the Eikonal equation, by considering such approximation. The application of Hamilton-Jacobi dynamics to ML is discussed in \cite{deepPDE}. 

\textbf{Possible applications: } One application of the SE-NET is design method for functional optical waveguides and lenses that enables end-to-end ML with physical structure; e.g. compressed sensing, computational imaging \cite{etoe1,etoe2,etoe3}, and optical communication \cite{huang2020demonstration}. Another application is an optical processor for ultrafast and energy-efficient inference engines \cite{YShen}. This is because the operation of the the transferred network is performed at the speed of light, which does not require any principal energy consumption

\textbf{Scalability of physical SE-NET: } Our SE-NET-based optical device requires less than 1 $\mu$m for each pixelized matrix, which is much smaller than in previous on-chip optical neural networks \cite{YShen, insitu}. This means that we can integrate over 100 million weights into 1-cm$^{2}$ optical chips, which is compatible with state-of-the-art application-specific integrated circuits (ASICs). In addition, we have not yet extracted the full potential of the SE-NET because there are many dimensions for the augmentation, including time, the $y$-axis, and polarization.

\textbf{From the viewpoint of ResNet: } Our model is inspired by the ResNet and its ODE representations. Although the ResNet has a very deep structure, its essence is an ensemble of relatively shallow neural networks, which was confirmed by the drop-out dependency of neural networks \cite{resnet1}. The SE-NET is formed by connecting each point as a node in the time direction. They can be seen as a superposition of many paths. The change in the path at a given point is obtained by multiplying the path by $exp(-jV(x,t)dt) \approx 1-jV(x,t)dt$, assuming that dt is very small. In this case, we get the unscattered path from the right-hand side first term and the scattered path from the right-hand side second term. Since dt is very small, the probability that a single path is scattered at many points is very small. The majority of the paths will be scattered by a small number. Furthermore, their superposition can be interpreted as an ensemble of relatively shallow networks. This is consistent with ResNet's analysis. This feature provides robustness against weight error, which is effective for the actual analog implementation described in Section 3. These investigations remain as future work.

\section{Related works}
\label{Related works}
\textbf{Neural ODEs: } The basic proposal of ODEs and their operation are described in section 2. Recently, network architectures based on ODE-Nets and their applications have been intensively investigated. For example, W. Grathwohl et al. proposed a neural ODE-based a generative model to solve inverse problems \cite{ffjord}. A higher-order approximation of the states in a neural ODE was proposed in \cite{journals/corr/abs-1905-10994}, and Y. Rubanova et al. proposed an ODE-based RNN structure \cite{journals/corr/abs-1907-03907}. The robustness of ODEs is discussed in \cite{Robust-ODE}. 

\textbf{Neural network with optics: } The crossover between neural networks and optics is a recent hot topic \cite{sui2020review}. On the application side, it has been reported that the joint optimization of the lens structure is effective for computational imaging applications \cite{etoe1,etoe2,etoe3}. However, the optimization is limited to a single-layer Fresnel lens, limiting the functionality of the optics. Furthermore, since  the optimization target is a point spread function (PSF),  the PSF has to be converted to an actual lens structure. By augmenting our method with a three-dimensional equation, we can optimize the local refractive index directly to enhance system performance.

On the basic research side, optical neuromorphic accelerators are being intensively investigated as candidates for  the DNN processor. They typically use a Mach-Zehnder interferometer \cite{YShen} or a discrete diffractive optical element \cite{lin2018all, lcos, meta} as a weight element. Adjoint optimization of these elements has already been shown in \cite{lin2018all,meta,insitu}. In general, increasing the number of nodes for each layer improves the characteristics in proportion to the polynomial of the number of nodes. On the other hand, increasing the number of layers is said to improve the characteristics exponentially, so it is better to have more layers. From this point of view, we think that the SE-NET is more advantageous than diffractive neural networks as a configuration to increase the number of layers. In addition, the SE-NET does not have a waveguide structure. Instead, each point becomes a node, and the nodes  are connected through the Hamiltonian. In contrast, an optical waveguides are difficult to scale up because the elements that optically connect them must be arranged in proportion to the square of the number of waveguides. 

\textbf{Complex-valued neural network:} Our proposed network acts as a complex-valued neural network. It has been demonstrated that the use of complex-valued parameters has numerous advantages. For example, Trabelsi et al. constructed the building blocks for complex-valued neural networks and demonstrated that the deep complex-valued network shows performance superior  to that of a real-valued one \cite{comdeep}. Arjovsky et al. have demonstrated that a recurrent neural network based on unitary weight matrices provides a richer representation \cite{uRNN}. 

\textbf{Quantum Machine Learning:} Our work is partially related to quantum ML because we use the Schr\"{o}dinger equation for ML. Among the various techniques, quantum circuit learning \cite{QCL1,QCL2}  is relatively related to our work. With this technique, the angular of unitary quantum  gates is trained to decrease the cost function through a backpropagation method, which is similar to our proposed WFM-based learning. Our approach differs from quantum circuit learning. We assume classical optical systems (not quantum systems), which are easier to fabricate and apply than quantum ones, but there is no quantum acceleration. In addition, we directly train the local potential field (not a gate device), allowing scalable system implementation. 

\section{Conclusion}
\label{conclusion}
We investigated the potential of Schr\"{o}dinger equation as a building block of neural networks. We show that the SE-NET can be trained by optimizing the local potential field of the Schr\"{o}dinger equation. As a demonstration, we trained the SE-NET by using the finite difference method and found its performance is comparable to the standard deep neural networks. The trained network is transferable to the actual optical systems. Our works extend the application field of machine learning from digital only optimization to hybrid physical digital optimizations. 

\clearpage

\section{Broader impact}
\label{Broader impact}
Our research bridges the gap between physical laws and neural networks. By augmenting the physical neural network concept, we believe that our work has the following impact. 

\textbf{Optimizing data accuision:} Recent machine learning (ML) techniques require complex processing and enormous datasets to achieve state-of-the-art performance. On the other hand, the input data are still captured to be easily understood by humans . It is questionable whether traditional data acquisition is the optimal approach for neural post-processing. For instance, it is difficult to get depth, spectral, and polarization information from standard commercial camera data. However, when we introduce special purpose lenses or gratings in front of sensors, we can easily reconstruct those data with simple processing. This implies that an optimized physical structure can simplify data processing and improve its performance. Our method provides a way to jointly optimize the physical structure and deep neural networks. Although the method has only been explored for simple tasks so far, it has the potential to integrate sensing and information processing in the future. 

\textbf{Physical structure as a computational resource:} The computational resources to execute deep learning are increasing more and more. Thus, alternative methods of computation are being widely explored to reduce energy consumption and processing latency. Based on our concept, we can outsource part of the digital processing to a passive physical structure, which will reduce the energy consumption of the whole processing operation. 

\textbf{Physics inspired ML} The crossover of ML and physics is a topic of great interest for, for instacne, material informatics and physics-inspired networks such as neural ODEs. Our paper shows the relationship between the manipulation of the physical quantityand the neural network, connecting ML and physics (details are described in the appendix). According to our idea, physicists can treat the physical equations as neural networks, and ML researchers can refer to physics for the design of DNNs .


\clearpage

\def\thesection{\Alph{section}}
\setcounter{section}{0}
\renewcommand{\theequation}{A.\arabic{equation} }
\setcounter{equation}{0}
\renewcommand{\thetable}{A.\arabic{table}}
\renewcommand{\thefigure}{A.\arabic{figure}}
\setcounter{figure}{0}
\setcounter{table}{0}

\section{Correspondence between physics and neural network}
\label{Correspondence between Optics, Mechanics, and Neural network}
We summarized the correspondence between physical phenomena and neural network in Figure A.1. The neural ODEs are the continuous representation of the ResNet or RNN\cite{conf/nips/ChenRBD18}. As discussed in the main article, Schr\"{o}dinger equation is considered as a continuous expression of complex-valued residual layer. Especially, it is a special type of unitary neural network \cite{uRNN} when we use the phase-only update rule described in the main article. Our discussions are directly connecting to waveguide optics and quantum mechanics because they are denoted by the Schr\"{o}dinger equation. The real-valued representation of the Schr\"{o}dinger equation corresponds to the diffusion equation \cite{okino2013}. The diffusion equation is typically used for modeling the thermal conduction and diffusion-reaction system. The discussion in section A.3 is effective by considering the real-valued case. Thus, diffusion equation is also considered as a special type of neural network. As is well known, classical limitation ($\hbar\rightarrow0$) in quantum systems, or paraxial approximation ($\lambda\rightarrow0$) in wave optics corresponds to the Hamilton-Jacobi equation, or Eikonal equation. Our work can be extended to analytical mechanics or analytical geometric optics. Note that the neural network based on such second-order ODE have already introduced by \cite{chang2018reversible}. The Schr\"{o}dinger equations derived from the scalar wave equation. Our work could thus be extended to complex-valued propagation of, for instance, electromagnetic waves and acoustic waves. The training of the such physical neural network can be executed by the adjoint method (reverse mode ODE or PDE) as the same manner as neural SE, which is essentially equivalent to the back propagation method used in the standard neural network. 

\begin{figure*}[ht]
\vskip 0.1in
\begin{center}
\centerline{\includegraphics[width=5 in]{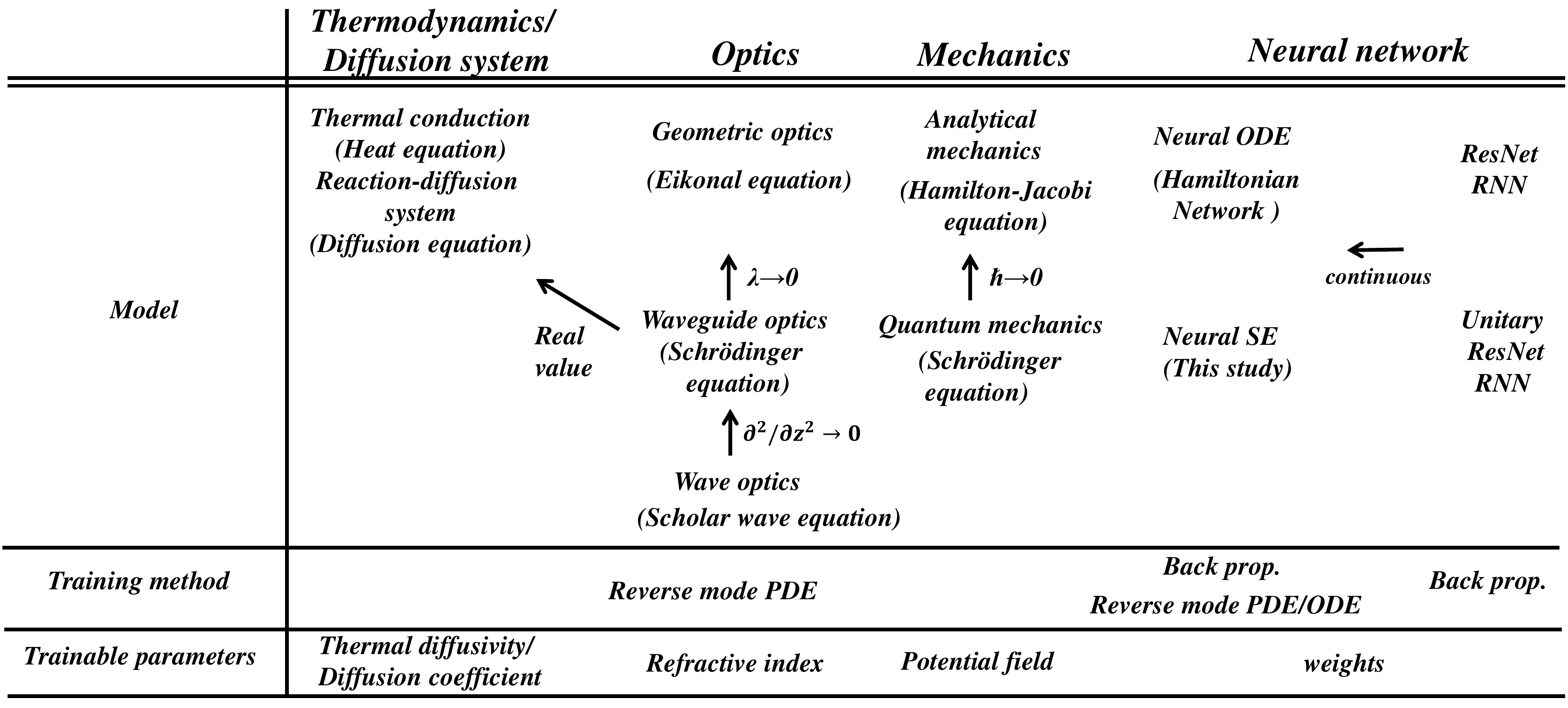}}
\caption{Correspondence between physical equation and neural networks. \cite{chang2018reversible}, \cite{uRNN}}
\label{icml-historical}
\end{center}
\vskip -0.4in
\end{figure*}

\begin{figure*}[t]
\vskip 0.1in
\begin{center}
\centerline{\includegraphics[width=6 in]{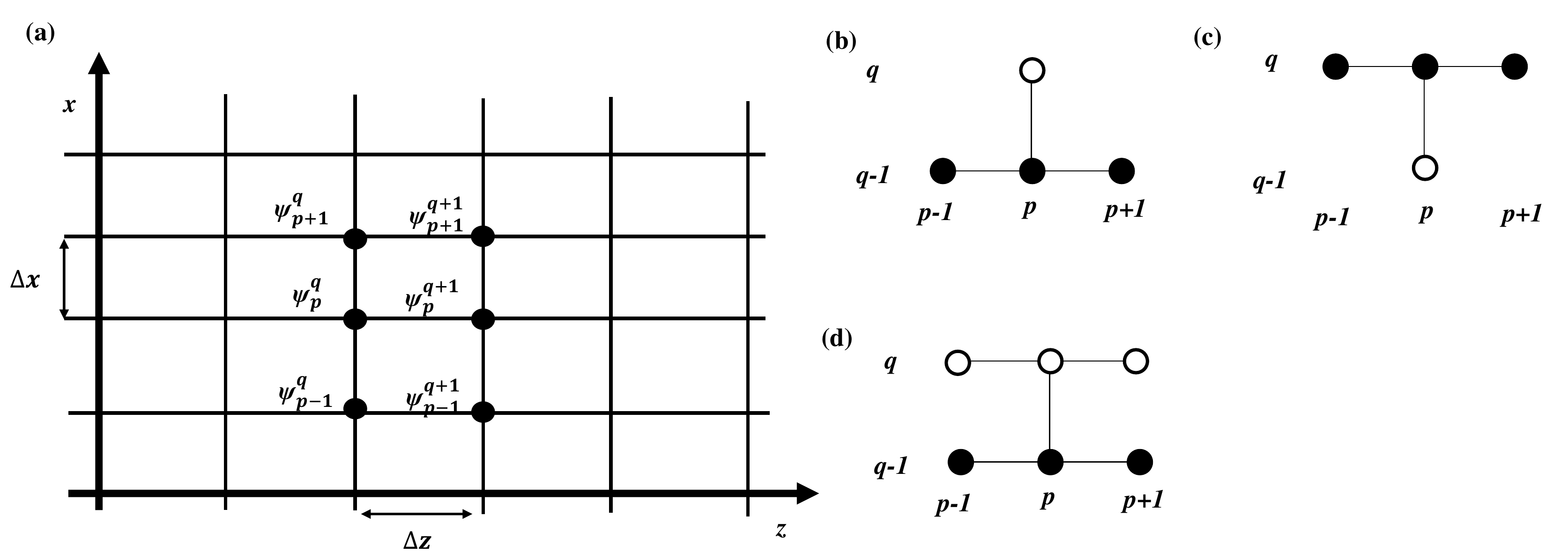}}
\caption{(a) Finite difference space grid. Diagram for computing FDM with (b) explicit ($\gamma=0$), (c) implicit ($\gamma=1$), and (d) Crank-Nicolson condition. $\bullet$ and $\circ$ denote known and unknown points}
\label{icml-historical}
\end{center}
\vskip -0.4in
\end{figure*}

\section{Crank-Nicolson Finite Difference Method}
\label{Crank-Nicolson Finite Difference Method}
We describe the Crank-Nicolson finite difference method (FDM) for solving the Schr\"{o}dinger equation numerically\cite{BPM_text}. In the FDM framework, an equation to solve is discretized by the linear spatial grid as shown in Fig. A.2(a). Then, by considering the first-order approximation: $\partial\psi/\partial z\approx(\psi_{p}^{q+1}-\psi_{p}^{q})/\Delta z$, Eq. (9) can be expressed as difference equation (A.1)

\begin{equation}j\frac{\psi_{p}^{q+1}-\psi_{p}^{q}}{\Delta z}=(1-\gamma)\mathcal{H}\psi_{p}^{q}+\gamma\mathcal{H}\psi_{p}^{q+1}.\end{equation}

where $\psi_{p}^{q}$ denotes the electric field at each grid position $(x, z) = (p\Delta x, q\Delta z)$, and the parameter $\gamma$ indicates the contribution of the forward (explicit) and backward (implicit) difference. For example, eq. (A.1) with $\gamma=0$ is known as an explicit method, and its calculation diagram is shown in Fig. A.2(b). This scheme is simple and easy to implement, but the solution becomes unstable when $\Delta x^{2}/\Delta z>1/2$. In addition, we need fine computational meshes to obtain the accurate solutions because their numerical error is relatively large $[O(\Delta x^{2}+\Delta z)]$. In the case of $\gamma=1$, eq. (A.1) becomes implicit method as shown in Fig. A.2(c). This scheme is numerically stable, but their numerical accuracy is same as that of the explicit method. On the other hand, Eq. (A.1) with $\gamma=1/2$ is known as the Crank-Nicolson method [Fig. A.2(d)]. This scheme is numerically stable and the numerical error is smaller than that of the explicit and implicit method $[O(\Delta x^{2}+\Delta z^{2})]$. Thus, Crank-Nicolson method is widely used. In this case, the Eq. (A.1) becomes
\begin{equation}j\frac{\psi_{p}^{q+1}-\psi_{p}^{q}}{\Delta z}=\mathcal{H}\frac{\psi_{p}^{q+1}-\psi_{p}^{q}}{2}.\end{equation}
By considering the definitions, 
\begin{equation} \mathcal{H_{L}}=\frac{1}{2kn_{r}}(-\frac{\partial^{2}}{\partial x^{2}}+k^2\Delta n_{p}^{q})\end{equation}
\begin{equation} \partial^{2}\psi/\partial z^{2}\approx\frac{\psi_{p+1}^{q}+\psi_{p-1}^{q}-2\psi_{p}^{q}}{\Delta z} \end{equation}
we can derive
\begin{equation} \begin{split} \psi_{p}^{q+1}&=\left(1+j\frac{1}{2}\mathcal{H}\Delta z\right)^{-1}\left(1-j\frac{1}{2}\mathcal{H}\Delta z\right)\psi_{p}^{q}\\
&\approx (u+D^{2}+v\Delta n_{p}^{q})^{-1} (u+D^{2}+v\Delta n_{p}^{q})\psi_{p}^{q}, \end{split}\end{equation}

where, $\Delta n_{p}^{q}$, $u$, and $v$ are defined as
\begin{equation}\Delta n_{p}^{q}=n(p\Delta x,q\Delta z)-n_{r}\end{equation}
\begin{equation} u=j\frac{4kn_{r}}{\Delta z}(\Delta x)^{2},\end{equation}
\begin{equation} v=2k^{2}n_{r}^{2}(\Delta x)^{2},\end{equation}

and $D$ is tridiagonal matrix described as follows;

\begin{equation}D= \left[\begin{array}{ccccccc}
-2&1&0&0&\ldots&0&0\\
1&-2&1&0&\ldots&0&0\\
0&1&-2&1&\ldots&0&0\\
0&0&1&-2&\ldots&0&0\\
\vdots&\vdots&\vdots&\vdots&\ddots&\vdots&\vdots\\
0&0&0&0&\ldots&-2&1\\
0&0&0&0&\ldots&1&-2\\
\end{array}\right],
\end{equation}
Thus we can compute the electric field by solving eq. (A.5) at each calculation grid by using linear solver. When we simply use a standard linear solver (e.g. Gaussian elimination, $numpy.linlg$ modules), $O(n_{x}^{3})$ computational resources are typically required. Here we use sparse matrix solver based on Thomas method. Then the computational resources decrease to $O(n_{x})$. The backward transmission can be solved in the same way. We implemented the above algorithm onto the Pytorch framework. Thus, we can train the SE-NET as a standard building block of deep neural networks.

\section{Convolutional neural network and PDE}
\label{Convolutional neural network and PDE}
We show that a convolution layer can be considered as a partial differential equation (PDE). The same discussion is shown in \cite{PDE-net}. Here we assume that one-dimensional PDE inputs $y \in \mathbb{C}^{n_{x}}$. It represents a one-dimensional grid function obtained at the cell centers with a mesh size $h= 1/n_{x}$. A convolution layer with one-dimensional Kernel filter is represented as $K \in \mathbb{C}^{(n_{x}\times n_{x})}$, which is parameterized by the convolution stencil $\theta \in \mathbb{C}^{3}$. Applying a coordinate change, the convolutional operation can be described as follows:
\begin{equation}\begin{split} K(\theta)y&=[\theta_{1}, \theta_{2}, \theta_{3}]\ast y\\
&=(\frac{\alpha_{1}}{4}[1,2,1]+\frac{\alpha_{2}}{2h}[-1,0,1]+\frac{\alpha_{3}}{h^{2}}[-1,2,-1])\ast y \end{split}\end{equation}
where $\ast$ denotes convolution operation, and $\alpha_{i}$ are given by the unique solution of (A.11).
\begin{equation} \left[\begin{array}{ccc}
1/4&-1/2h&-1/h^{2}\\
1/2&0&2/h^{2}\\
1/4&1/2h&-1/h^{2}\end{array}\right]
\left[\begin{array}{ccc}
\alpha_{1}\\
\alpha_{2}\\
\alpha_{3}\end{array}\right]
=
\left[\begin{array}{ccc}
\theta_{1}\\
\theta_{2}\\
\theta_{3}
\end{array}\right]
\end{equation}

By considering the limitation $h\rightarrow0$, 

\begin{equation} K(\theta)=\alpha_{1}(\theta)+\frac{\partial}{\partial x}\alpha_{2}(\theta)+\frac{\partial^{2}}{\partial x^{2}}\alpha_{3}(\theta).\end{equation}

This argument extends to higher spatial dimensions and it acts as a high-dimensional Kernel filter.

\section{End-to-end learning method}
\label{etoe}
As a example of end-to-end ML, we consider the system for the spectroscopy as shown in Fig. 4 in the main article. The detailed processing scheme is shown in Fig. A.3. It consists of two-dimensional optical circuits that act as SE-NETs, photodetectors (PDs) to receive the SE-NET outputs, and digital electric circuits which act as a standard neural network. The input power spectra $u(\lambda)$ are represented as a complex-valued Gaussian beam, which described as follows.
\begin{equation}\psi(x,z_0,\lambda) = \sqrt{u(\lambda)}\Phi (x)\end{equation}
where $\Phi$ represents the input waveguide fields as follows;
\begin{equation} \Phi(x)=\frac{1}{\sqrt{2\pi\omega_{i}^{2}}}\exp \left[-\frac{x^{2}}{\omega_{i}^{2}}\right]\end{equation}
where $\omega_{i}$ is mode radius of the input edge of optimizing regeion. The data is processed by SE-NETs (Eq. 9) and detected by the PD. The detection efficiency $\eta$ of each PD for each wavelength $\lambda$ is described as
\begin{equation} \eta_{i}(\lambda)=\int \psi(x,z_1,\lambda)\Phi_{i}'(x)dx\end{equation}
where $i$ denotes the detector number, and $\Phi'$ represents the detection fields of PDs, which are typically described as Gaussian profiles.
\begin{equation} \Phi_{i}'(x)=\frac{1}{\sqrt{2\pi\omega_{o}^{2}}}\exp \left[-\frac{(x-ix_{p})^{2}}{\omega_{o}^{2}}\right]\end{equation}
where $\omega_{o}$ is the mode radius of output wavegide, and $x_{p}$ is PD channel spacing. The total output power is described as 
\begin{equation} X_{i}=\int |\eta_{i}(\lambda)|^{2}d\lambda\end{equation}
The detected data are acquired by digital circuits, and they become inputs of digital DNNs. DNN output $Y$ is converted to loss $L$ by the cost function. Then, the backward of $L$ is computed by using the standard backward process, and we get following equations:
\begin{equation} \frac{\partial L}{\partial X_{i}}=(\frac{\partial Y}{\partial X_{i}})(\frac{\partial \eta}{\partial Y})\end{equation}
\begin{equation} \psi_{a}(x,z_{1})=\sum_{i}^{M}2\Phi_{i}'(x)Re[\eta_{i}(\lambda)]\frac{\partial L}{\partial X_{i}}\end{equation}
where $M$ is the number of PDs. As the backward in analog photonics $\psi_{a}(x,z)$ is solved above equation, we can design the optimal structure through the end-to-end hybrid machine learning. 

\begin{figure}[t]
\vskip 0.1in
\begin{center}
\centerline{\includegraphics[width=\columnwidth]{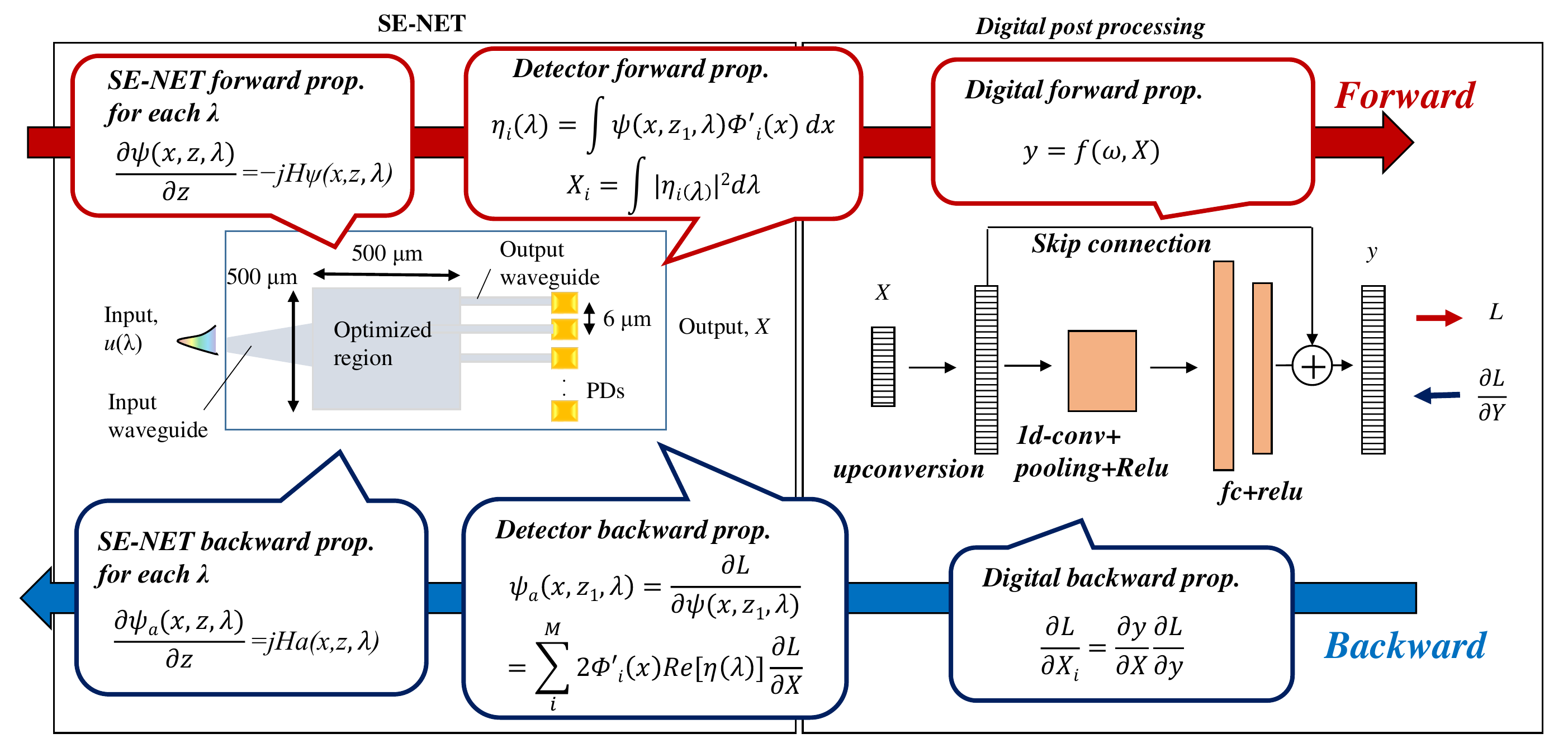}}
\caption{ End-to-end machine learning with optical frontend}
\label{icml-historical}
\end{center}
\vskip 0in
\end{figure}

For the experiment in Fig. 4 in the main article, we set the parameters as follows. The first layer is SE-NET using the optical waveguide. The $\omega_{i} $ and $\omega_{o}$ are set to 100 and 6 $\mu$m, respectively. The analyzed region was set to $500\times500 \mu$m$^{2}$. The 25 detectors were set on the output edge of the waveguide (M=25) with $x_{p} = 12 \mu m$. We used a linear Hamiltonian by considering easy transfer to the optics. For the post processing, we used a residual layer with a one-dimensional convolution filter, which has almost the same structure as the filter in \cite{CS2}. It consists the up-conversion layer, one-dimensional convolution layer, two fully-connected (fc) layers, and a residual connection layer. The kernel size and padding size of the convolution filter are set to 16 and 8. For the up-conversion, we use the pseudo-inverse matrix of the transmission matrix $T$.

\begin{figure}[t]
\vskip 0.1in
\begin{center}
\centerline{\includegraphics[width=10cm]{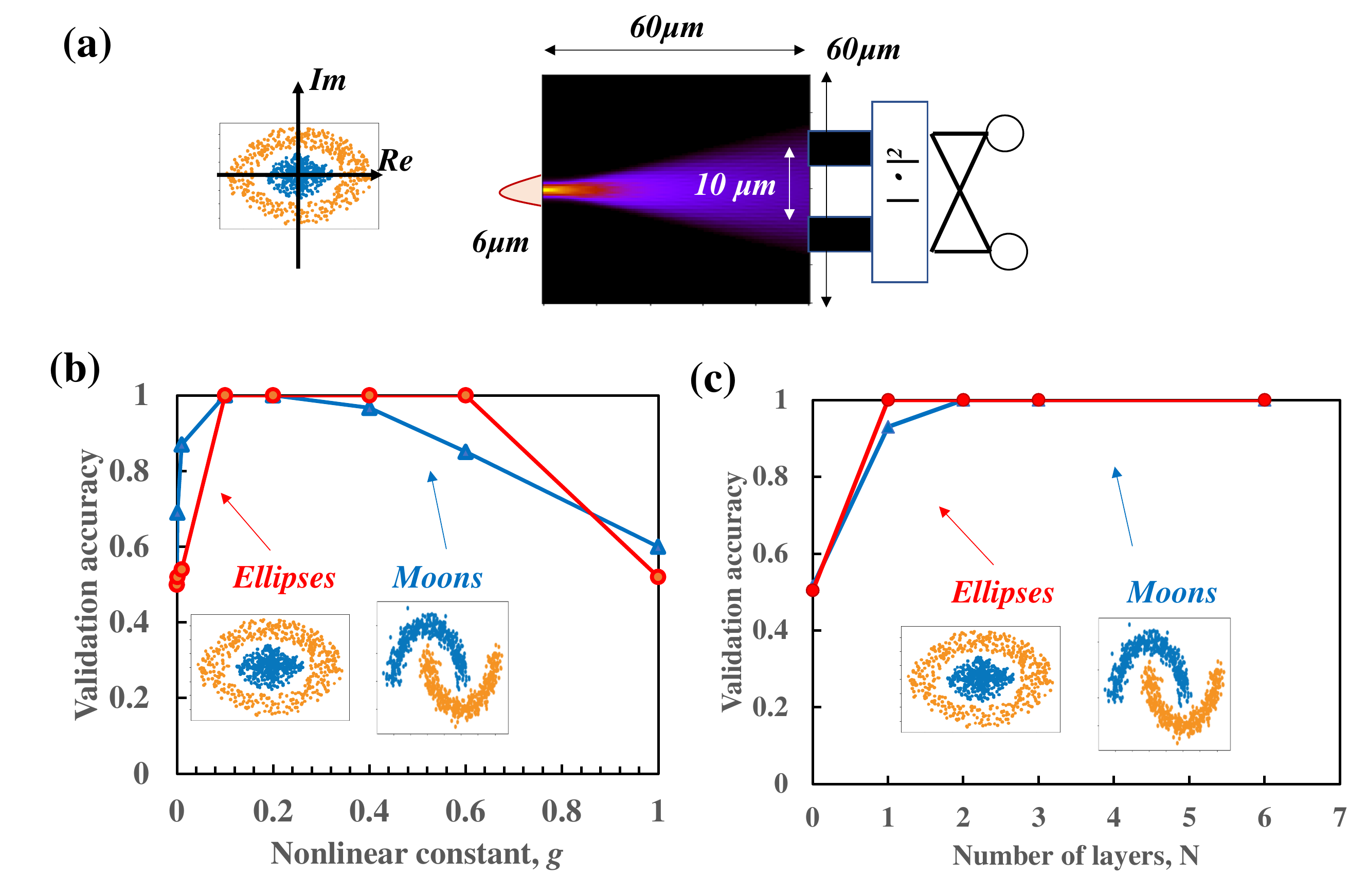}}
\caption{(a) Simulation set-up. Results for the SE-NET with (b) nonlinear Hamiltonian and (c) cascade connected networks.}
\label{icml-historical}
\end{center}
\vskip -0.4in
\end{figure}
\section{Additional Experiments}

\textbf{Binary classification task: } We consider a small-scale test problem in two dimensions called ellipses and moons. The experimental set-up is shown in Fig. A.4(a).The test data consists of 1,200 points that are evenly divided into two groups as shown in the subplots in Figure A.4(b) and (c). The original data is randomly divided into 1,000 training examples and 200 examples used for validation. We train the SE-NETs with the nonlinear Hamiltonian [Fig. 3(c)] and Cascade connected network [Fig. 3(d)]. For the activation function of the cascade connected network, the tanh function was employed, which can be physically implemented by using the saturable absorbers. The input data are represented as a complex-valued Gaussian beam with a beam width of 6 $\mu$m. The analyzed region was set to $60\times60 \mu$m$^{2}$. The nonlinearity of the networks was scanned by changing the $g$ value (for the nonlinear Hamiltonian case) and divided layer number $N$ (for the cascaded network case). At the output edge, the intensity was detected by two-arrayed PDs with $\omega_{o}=6\mu$m. We show the results in Figure A3(b) and (c). As the examined task obviously requires nonlinear conversion, the linear SE-NETs [$g=0$ in Fig. A.4(b), $N=0$ in Fig. A.4(c)] could not classify the data. When the network have a nonlinearity, the validation accuracy increases up to 100\%. When $g$ increase too much, the accuracy was degraded. Thus, we set $g=0.1$ for another experiment. 

\textbf{CIFAR10: } We trained SE-NETs to classify the CIFAR10 image dataset. The data set consists of 60,000 labeled digital images with size $32\times32$ . For the training, the optimizer steps were computed using mini-batches consisting of 128 randomly chosen examples from the 60,000 training data. The input wavelength was set to $(\lambda=2\pi/k)$ to 1.55 $\mu $m. The trained models is almost same as CNN-SE-NET in Fig. 3(a). The differences are follows; (i) the input dimension of the convolution filter was set to three to match the data dimension of CIFAR10, (ii) the region of SE-NET was set to $1150 \times 400 \mu m^{2}$. It has a CNN-based down sampling layer containing three two-dimensional convolution units, which is the same as in the pre-filtering method used in the previously reported ODE-Net model \cite{conf/nips/ChenRBD18}. We consider the nonlinear activation model as shown in Fig. 3(d). This system can be physically devised by inserting the nonlinear material discretely. At the input edge of the SE-NETs, the input data are aligned to the $x$-axis and mapped to the real part of $\psi(x,z_{0})$. At the output edge of SE-NETs, the real and imaginary parts of $\psi(x,z=z_{1})$ are independently treated. They go to the full connection layer. The experimental results for test accuracy was 71.4 \% after the 30-epoch training.

\bibliographystyle{acm}
\bibliography{ref3}

\begin{thebibliography}{10}

\bibitem{uRNN}
{\sc Arjovsky, M., Shah, A., and Bengio, Y.}
\newblock Unitary evolution recurrent neural networks.
\newblock In {\em International Conference on Machine Learning\/} (2016),
  pp.~1120--1128.

\bibitem{CS4}
{\sc August, Y., and Stern, A.}
\newblock Compressive sensing spectrometry based on liquid crystal devices.
\newblock {\em Optics letters 38}, 23 (2013), 4996--4999.

\bibitem{meta}
{\sc Backer, A.~S.}
\newblock Computational inverse design for cascaded systems of metasurface
  optics.
\newblock {\em Optics express 27}, 21 (2019), 30308--30331.

\bibitem{chang2018reversible}
{\sc Chang, B., Meng, L., Haber, E., Ruthotto, L., Begert, D., and Holtham, E.}
\newblock Reversible architectures for arbitrarily deep residual neural
  networks.
\newblock In {\em Thirty-Second AAAI Conference on Artificial Intelligence\/}
  (2018).

\bibitem{lcos}
{\sc Chang, J., Sitzmann, V., Dun, X., Heidrich, W., and Wetzstein, G.}
\newblock Hybrid optical-electronic convolutional neural networks with
  optimized diffractive optics for image classification.
\newblock {\em Scientific reports 8}, 1 (2018), 1--10.

\bibitem{etoe3}
{\sc Chang, J., and Wetzstein, G.}
\newblock Deep optics for monocular depth estimation and 3d object detection.
\newblock In {\em Proceedings of the IEEE International Conference on Computer
  Vision\/} (2019), pp.~10193--10202.

\bibitem{chang1999difference}
{\sc Chang, Q., Jia, E., and Sun, W.}
\newblock Difference schemes for solving the generalized nonlinear
  schr{\"o}dinger equation.
\newblock {\em Journal of Computational Physics 148}, 2 (1999), 397--415.

\bibitem{deepPDE}
{\sc Chaudhari, P., Oberman, A., Osher, S., Soatto, S., and Carlier, G.}
\newblock Deep relaxation: partial differential equations for optimizing deep
  neural networks.
\newblock {\em Research in the Mathematical Sciences 5}, 3 (2018), 30.

\bibitem{conf/nips/ChenRBD18}
{\sc Chen, T.~Q., Rubanova, Y., Bettencourt, J., and Duvenaud, D.}
\newblock Neural ordinary differential equations.
\newblock In {\em NeurIPS}, S.~Bengio, H.~M. Wallach, H.~Larochelle,
  K.~Grauman, N.~Cesa-Bianchi, and R.~Garnett, Eds., pp.~6572--6583.

\bibitem{soloton}
{\sc Eisenberg, H.~S., Silberberg, Y., Morandotti, R., Boyd, A.~R., and
  Aitchison, J.~S.}
\newblock Discrete spatial optical solitons in waveguide arrays.
\newblock {\em Phys. Rev. Lett. 81}, 16 (Oct. 1998), 3383--3386.

\bibitem{etoe2}
{\sc Elmalem, S., Giryes, R., and Marom, E.}
\newblock Learned phase coded aperture for the benefit of depth of field
  extension.
\newblock {\em Optics express 26}, 12 (2018), 15316--15331.

\bibitem{ffjord}
{\sc Grathwohl, W., Chen, R.~T., Bettencourt, J., Sutskever, I., and Duvenaud,
  D.}
\newblock Ffjord: Free-form continuous dynamics for scalable reversible
  generative models.
\newblock {\em arXiv preprint arXiv:1810.01367\/} (2018).

\bibitem{graves2013speech}
{\sc Graves, A., Mohamed, A.-r., and Hinton, G.}
\newblock In {\em 2013 IEEE International Conference on Acoustics, Speech and
  Signal Processing}.

\bibitem{WFM1}
{\sc Hashimoto, T., Saida, T., Ogawa, I., Kohtoku, M., Shibata, T., and
  Takahashi, H.}
\newblock Optical circuit design based on a wavefront-matching method.
\newblock {\em Optics letters 30}, 19 (2005), 2620--2622.

\bibitem{wave-theory}
{\sc Haus, H.~A.}
\newblock {\em Waves and fields in optoelectronics}.
\newblock Prentice-Hall,, 1984.

\bibitem{QCL2}
{\sc Havl{\'\i}{\v{c}}ek, V., C{\'o}rcoles, A.~D., Temme, K., Harrow, A.~W.,
  Kandala, A., Chow, J.~M., and Gambetta, J.~M.}
\newblock Supervised learning with quantum-enhanced feature spaces.
\newblock {\em Nature 567}, 7747 (2019), 209--212.

\bibitem{CS3}
{\sc Heiser, Y., Oiknine, Y., and Stern, A.}
\newblock Compressive hyperspectral image reconstruction with deep neural
  networks.
\newblock In {\em Big Data: Learning, Analytics, and Applications\/} (2019),
  vol.~10989, International Society for Optics and Photonics, p.~109890M.

\bibitem{bp-rc}
{\sc Hermans, M., Antonik, P., Haelterman, M., and Massar, S.}
\newblock Embodiment of learning in electro-optical signal processors.
\newblock {\em Physical review letters 117}, 12 (2016), 128301.

\bibitem{huang2020demonstration}
{\sc Huang, C., Fujisawa, S., de~Lima, T.~F., Tait, A.~N., Blow, E., Tian, Y.,
  Bilodeau, S., Jha, A., Yaman, F., Batshon, H.~G., et~al.}
\newblock Demonstration of photonic neural network for fiber nonlinearity
  compensation in long-haul transmission systems.
\newblock In {\em 2020 Optical Fiber Communications Conference and Exhibition
  (OFC)\/} (2020), IEEE, pp.~1--3.

\bibitem{insitu}
{\sc Hughes, T.~W., Minkov, M., Shi, Y., and Fan, S.}
\newblock Training of photonic neural networks through in situ backpropagation
  and gradient measurement.
\newblock {\em Optica 5}, 7 (2018), 864--871.

\bibitem{adjointsip}
{\sc Hughes, T.~W., Minkov, M., Williamson, I.~A., and Fan, S.}
\newblock Adjoint method and inverse design for nonlinear nanophotonic devices.
\newblock {\em ACS Photonics 5}, 12 (2018), 4781--4787.

\bibitem{CS1}
{\sc Kim, C., Park, D., and Lee, H.-N.}
\newblock Convolutional neural networks for the reconstruction of spectra in
  compressive sensing spectrometers.
\newblock In {\em Optical Data Science II\/} (2019), vol.~10937, International
  Society for Optics and Photonics, p.~109370L.

\bibitem{CS2}
{\sc Kim, C., Park, D., and Lee, H.-N.}
\newblock Compressive sensing spectroscopy using a residual convolutional
  neural network.
\newblock {\em Sensors 20}, 3 (2020), 594.

\bibitem{lecun2015deeplearning}
{\sc LeCun, Y., Bengio, Y., and Hinton, G.}
\newblock Deep learning.
\newblock {\em nature 521}, 7553 (2015), 436--444.

\bibitem{lenet1}
{\sc LeCun, Y., Bottou, L., Bengio, Y., and Haffner, P.}
\newblock Gradient-based learning applied to document recognition.
\newblock {\em Proceedings of the IEEE 86}, 11 (1998), 2278--2324.

\bibitem{MNIST}
{\sc LeCun, Y., Cortes, C., and Burges, C.~J.}
\newblock The mnist database of handwritten digits, 1998.
\newblock {\em URL http://yann. lecun. com/exdb/mnist 10\/} (1998), 34.

\bibitem{lecun1}
{\sc LeCun, Y., Touresky, D., Hinton, G., and Sejnowski, T.}
\newblock A theoretical framework for back-propagation.
\newblock In {\em Proceedings of the 1988 connectionist models summer school\/}
  (1988), vol.~1, CMU, Pittsburgh, Pa: Morgan Kaufmann, pp.~21--28.

\bibitem{journals/corr/LillicrapHPHETS15}
{\sc Lillicrap, T.~P., Hunt, J.~J., Pritzel, A., Heess, N., Erez, T., Tassa,
  Y., Silver, D., and Wierstra, D.}
\newblock Continuous control with deep reinforcement learning.
\newblock In {\em ICLR}, Y.~Bengio and Y.~LeCun, Eds.

\bibitem{CS5}
{\sc Lin, X., Liu, Y., Wu, J., and Dai, Q.}
\newblock Spatial-spectral encoded compressive hyperspectral imaging.
\newblock {\em ACM Transactions on Graphics (TOG) 33}, 6 (2014), 1--11.

\bibitem{lin2018all}
{\sc Lin, X., Rivenson, Y., Yardimci, N.~T., Veli, M., Luo, Y., Jarrahi, M.,
  and Ozcan, A.}
\newblock All-optical machine learning using diffractive deep neural networks.
\newblock {\em Science 361}, 6406 (2018), 1004--1008.

\bibitem{QCL1}
{\sc Mitarai, K., Negoro, M., Kitagawa, M., and Fujii, K.}
\newblock Quantum circuit learning.
\newblock {\em Physical Review A 98}, 3 (2018), 032309.

\bibitem{okino2013}
{\sc Okino, T.}
\newblock Correlation between diffusion equation and schr{\"o}dinger equation.

\bibitem{BPM_text}
{\sc Pedrola, G.~L.}
\newblock {\em Beam Propagation Method for Design of Optical Waveguide
  Devices}.
\newblock John Wiley \& Sons, 2015.

\bibitem{journals/corr/abs-1907-03907}
{\sc Rubanova, Y., Chen, R. T.~Q., and Duvenaud, D.}

\bibitem{PDE-net}
{\sc Ruthotto, L., and Haber, E.}
\newblock Deep neural networks motivated by partial differential equations.
\newblock {\em CoRR\/}.

\bibitem{WFM2}
{\sc Sakamaki, Y., Saida, T., Hashimoto, T., and Takahashi, H.}
\newblock New optical waveguide design based on wavefront matching method.
\newblock {\em Journal of lightwave technology 25}, 11 (2007), 3511--3518.

\bibitem{YShen}
{\sc Shen, Y., Harris, N.~C., Skirlo, S., Prabhu, M., Baehr-Jones, T.,
  Hochberg, M., Sun, X., Zhao, S., Larochelle, H., Englund, D., et~al.}
\newblock Deep learning with coherent nanophotonic circuits.
\newblock {\em Nature Photonics}, 7, 441.

\bibitem{silver2016mastering}
{\sc Silver, D., Huang, A., Maddison, C.~J., Guez, A., Sifre, L., van~den
  Driessche, G., Schrittwieser, J., Antonoglou, I., Panneershelvam, V.,
  Lanctot, M., Dieleman, S., Grewe, D., Nham, J., Kalchbrenner, N., Sutskever,
  I., Lillicrap, T., Leach, M., Kavukcuoglu, K., Graepel, T., and Hassabis, D.}
\newblock Mastering the game of go with deep neural networks and tree search.
\newblock {\em Nature\/}, 484--.

\bibitem{etoe1}
{\sc Sitzmann, V., Diamond, S., Peng, Y., Dun, X., Boyd, S., Heidrich, W.,
  Heide, F., and Wetzstein, G.}
\newblock End-to-end optimization of optics and image processing for achromatic
  extended depth of field and super-resolution imaging.
\newblock {\em ACM Transactions on Graphics (TOG) 37}, 4 (2018), 1--13.

\bibitem{sui2020review}
{\sc Sui, X., Wu, Q., Liu, J., Chen, Q., and Gu, G.}
\newblock A review of optical neural networks.
\newblock {\em IEEE Access 8\/} (2020), 70773--70783.

\bibitem{comdeep}
{\sc Trabelsi, C., Bilaniuk, O., Zhang, Y., Serdyuk, D., Subramanian, S.,
  Santos, J.~F., Mehri, S., Rostamzadeh, N., Bengio, Y., and Pal, C.~J.}
\newblock Deep complex networks.
\newblock {\em arXiv preprint arXiv:1705.09792\/} (2017).

\bibitem{resnet1}
{\sc Veit, A., Wilber, M.~J., and Belongie, S.}
\newblock Residual networks behave like ensembles of relatively shallow
  networks.
\newblock In {\em Advances in neural information processing systems\/} (2016),
  pp.~550--558.

\bibitem{Robust-ODE}
{\sc YAN, H., DU, J., TAN, V., and FENG, J.}
\newblock On robustness of neural ordinary differential equations.
\newblock In {\em International Conference on Learning Representations}.

\bibitem{journals/corr/abs-1905-10994}
{\sc Yildiz, C., Heinonen, M., and Lahdesmaki, H.}
\newblock Ode$^2$vae: Deep generative second order odes with bayesian neural
  networks.
\newblock {\em CoRR\/}.

\end{thebibliography}

\end{document}